\documentclass[11pt]{article}
\usepackage{latexsym,cite,amssymb,amsmath}

\usepackage{times}
\usepackage{hyperref} 
\makeatletter
\@addtoreset{equation}{section} \makeatother

\input amssym.def
\input amssym.tex

\newcommand{\half}{{{\textstyle\frac{1}{2}}}}

\newcommand{\be}{\begin{equation}}
\newcommand{\ee}{\end{equation} }
\newcommand{\beqa}{\begin{eqnarray} }
\newcommand{\eeqa}{\end{eqnarray} }
\newcommand{\ba}{\begin{array}}
\newcommand{\ea}{\end{array}}

\newcommand{\ODD}{\mathbf{O}(D,D)}

\newcommand{\mbg}{\mathbf{g}}

\newcommand{\G}{\mathbf{G}}
\newcommand{\na}{{\nabla}}

\newcommand\Tr{{\rm Tr}}

\newcommand\rd{{\rm d}}

\newcommand\eff{{\rm eff.}}

\newcommand\cD{{\cal D}}

\newcommand\cF{{\cal F}}
\newcommand\cG{{\cal G}}
\newcommand\cH{{\cal H}}

\newcommand\cR{{\cal R}}

\newcommand\hcF{{\hat{\cal F}}}

\newcommand\tcL{{\widetilde{\cal L}}}

\newcommand\rbC{{\scriptscriptstyle{\mathbf{C}}}}

\newcommand\gYM{g_{\rm \scriptscriptstyle{YM}}}

\newcommand\dis{\displaystyle}

\def\tx{\tilde{x}}

\def\Tw{{T}}

\def\bP{{\bar{P}}}

\begin{document}
\begin{titlepage}
\title{%\vskip -60pt
%\vskip 20pt
\vskip 2cm
Double field formulation  of Yang-Mills theory \\~\\}
\author{\sc Imtak Jeon,${}^{\dagger}$  \mbox{~~\,\,}
Kanghoon Lee${}^{\sharp}$ \mbox{~~\,\,}and\mbox{\,\,~~} Jeong-Hyuck Park${}^{\dagger}$}
\date{}
\maketitle \vspace{-1.0cm}
\begin{center}
~~~\\
${}^{\sharp}$Center for Quantum Spacetime, Sogang University, Shinsu-dong, Mapo-gu, Seoul 121-742, Korea\\
%\texttt{kanghoon@sogang.ac.kr}
%\texttt{}\\
~{}\\
${}^{\dagger}$Department of Physics, Sogang University, Shinsu-dong, Mapo-gu, Seoul 121-742, Korea\\
%\texttt{imtak@sogang.ac.kr    ~~~~~park@sogang.ac.kr}
~{}\\
%{\small{{Electronic correspondence:} \texttt{{{{~imtak,kanghoon,park@sogang.ac.kr}}}}}}\\
{\small{{E-mail:} \texttt{~{{{imtak@sogang.ac.kr~~kanghoon@sogang.ac.kr~~park@sogang.ac.kr}}}}}}
~~~\\~\\~\\
\end{center}
\begin{abstract}
\vskip0.2cm
\noindent
Based on our previous work on the differential geometry for  the closed string  double field theory,  
 we construct a Yang-Mills action  which is covariant under $\ODD$ T-duality  rotation and invariant under 
three-types of  gauge transformations: non-Abelian Yang-Mills, diffeomorphism and one-form gauge symmetries.  In double field formulation, in a manifestly covariant manner   our action couples a single   $\ODD$ vector potential to the  closed string  double field theory.  In terms of  undoubled  component fields, it   couples  a usual  Yang-Mills gauge field to an additional  one-form field   and also to the closed string background fields which consist of  a dilaton,  graviton and a two-form gauge field.   Our resulting action resembles a twisted Yang-Mills action.    
 \end{abstract}

%%%
{\small
\begin{flushleft}
~~~~~~~~\textit{PACS}: 11.25.-w\\
%%Strings and branes \\
~~~~~~~~\textit{Keywords}: Double field theory, T-duality, Differential geometry.
\end{flushleft}}
\thispagestyle{empty}
%%%%
\end{titlepage}
\newpage
\tableofcontents %%
%%\begin{document} --> JHEP
%%%
%%%%%%%%%%%%%%%%%%%%%%%%%%%%%%%%%%%%%%%%%%%%%%%%%%%%%%%%%%%%%%%%%%%%%%%%%%%%%%%%%%%%%%%%%%%%%%%%
%%%%%%%%%%%%%%%%%%%%%%%%%%%%%%%%%%%%%%%%%%%%%%%%%%%%%%%%%%%%%%%%%%%%%%%%%%%%%%%%%%%%%%%%%%%%%%%%
{}~\\
~\\
~\\
\section{Introduction}
The  low energy effective action for a closed string massless sector takes the following well-known  form:
\be
\dis{S_{\eff}=\int\rd x^{D}\sqrt{-g}e^{-2\phi}\left[\,R_{g}+4(\partial \phi)^{2}-\textstyle{\frac{1}{12}}H^{2}\,\right]\,,}
\label{NSaction}
\ee
where $g_{\mu\nu}$ is the  $D$-dimensional   spacetime metric with its scalar curvature, $R_{g\,}$;    $\phi$ is the  string theory dilaton; and $H$ is the three form field strength of a two form gauge field, $B_{\mu\nu}$.  In  a double field theory (DFT) formalism developed by Hull  \textit{et all,} in \cite{Hull:2009mi,Hull:2009zb,Hohm:2010jy,Hohm:2010pp},  the above action was  reformulated as
%%%
%%\be
%%\ba{ll}
%%\multicolumn{2}{c}{\dis{S_{\scriptstyle{\rm{DFT}}}=\int\rd y^{2D}~e^{-2d}\,\cR(\cH,d)\,,}}\\
%%\cR(\cH,d)=&\cH^{AB}\left(4\partial_{A}\partial_{B}d-4\partial_{A}d\partial_{B}d+\textstyle{\frac{1}{8}}
%%\partial_{A}\cH^{CD}\partial_{B}\cH_{CD}
%%-\half\partial_{A}\cH^{CD}\partial_{C}\cH_{BD}\right)\\
%%{}&~+4\partial_{A}\cH^{AB}\partial_{B}d-\partial_{A}\partial_{B}\cH^{AB}\,.
%%\ea
%%\label{DFTS}
%%\ee
%%%
\be
\ba{ll}
\dis{S_{\scriptstyle{\rm{DFT}}}=\int\rd y^{2D}~e^{-2d}\Big[}&\!\!\!\cH^{AB}\left(4\partial_{A}\partial_{B}d-4\partial_{A}d\partial_{B}d+\textstyle{\frac{1}{8}}
\partial_{A}\cH^{CD}\partial_{B}\cH_{CD}
-\half\partial_{A}\cH^{CD}\partial_{C}\cH_{BD}\right)\\
{}&+4\partial_{A}\cH^{AB}\partial_{B}d-\partial_{A}\partial_{B}\cH^{AB}~\Big]\,.
\ea
\label{DFTS}
\ee
Herein  the spacetime dimension is formally doubled from $D$ to $2D$ with coordinates  $x^{\mu}\rightarrow y^{A}=(\tx_{\mu},x^{\nu})$; $d$ denotes   the double field theory `dilaton' given by $e^{-2d}=\sqrt{-g}e^{-2\phi}$; and  $\cH_{AB}$ is  a $2D\times 2D$ matrix of the form,
\be
\cH_{AB}=\left(\ba{cc}
g^{\mu\nu}&-g^{\mu\kappa}B_{\kappa\sigma}\\
B_{\rho\kappa}g^{\kappa\nu}&~~g_{\rho\sigma}-B_{\rho\kappa}g^{\kappa\lambda}B_{\lambda\sigma}
\ea
\right)\,.
\label{gH}
\ee
All the spacetime indices, $A,B,C,\cdots,$ are $2D$-dimensional vector  indices which can be  raised or lowered by the   $\ODD$ invariant constant metric, $\eta$, 
\be
{\eta:={\textstyle{\left(\ba{cc}0&1\\1&0\ea\right)}}\,.}
\label{ODDeta}
\ee
As a  field theory counterpart of the level matching condition in  closed string theory,  it is required that,\footnote{Note that throughout our paper, the equivalence symbol, `$\equiv$', denotes the equality up to the level matching  constraint (\ref{constraint}).}  
all the  fields  in  double field theory as well as  all of their possible products should be  annihilated by the $\ODD$ d'Alembert operator, $\partial^{2}=\partial_{A}\partial^{A}$,
\be
\ba{ll}
\partial^{2}\Phi\equiv 0\,,~~~~&~~~~
\partial_{A}\Phi_{1}\partial^{A}\Phi_{2}\equiv 0\,.
\ea
\label{constraint}
\ee
This constraint, which one may call `the level matching constraint',  actually   means  that  the theory is not truly doubled:  there is a choice of coordinates $(\tx^{\prime},x^{\prime})$, related to the original coordinates $(\tx,x)$, by an  
$\ODD$ rotation, in which all the  fields do not depend on the $\tx^{\prime}$ coordinates~\cite{Hohm:2010jy}. Remarkably, while   the   double field theory action, $S_{\scriptstyle{\rm{DFT}}}$  (\ref{DFTS}), reduces to the effective action,  $S_{\eff}$ (\ref{NSaction}), upon the level matching  constraint,  the  double field theory formulation  manifests the $\ODD$ covariance of the action\footnote{Without  imposing the  level matching  constraint, the $\ODD$ transformation   surely corresponds to a Noether  symmetry of the $2D$-dimensional  field theory. After imposing the constraint, the double field theory is, by nature,  $D$-dimensional: it  lives  on a  $D$-dimensional hyperplane. As the $\ODD$ transformation then  rotates the entire  hyperplane,  the $\ODD$ rotation acts  \textit{a priori} as  a duality   rather than a Noether symmetry of the $D$-dimensional theory.   After  further dimensional reductions,  it becomes a Noether symmetry of the reduced action,  as verified  by Buscher~\cite{Buscher:1985kb,Buscher:1987sk,Buscher:1987qj} (\textit{c.f.~} \cite{Hohm:2011gs}).} and hence    the T-duality   first noted by Buscher~\cite{Buscher:1985kb,Buscher:1987sk,Buscher:1987qj} and further studied  in \cite{Giveon:1988tt,Tseytlin:1990nb,Tseytlin:1990va,Siegel:1993th,Siegel:1993xq,Alvarez:1994wj,Giveon:1994fu,Grana:2008yw}. \\

However, what is not obvious  about  the  above DFT action (\ref{DFTS}) is that it possesses   gauge symmetry, which must be the case~\cite{Hohm:2010pp,Kwak:2010ew},  since restricted on the $x$-hyperplane the action (\ref{DFTS}) is nothing but  a rewriting of the effective action (\ref{NSaction}) while  the latter surely  enjoys both the $D$-dimensional diffeomorphism, $x^{\mu}\rightarrow x^{\mu}+\delta x^{\mu}$, and the gauge symmetry of the two form field, $B_{\mu\nu}\rightarrow B_{\mu\nu}+\partial_{\mu}\Lambda_{\nu} - \partial_{\nu}\Lambda_{\mu}$.  That is to say,  in contrast to the effective action (\ref{NSaction}) where  the gauge symmetry is manifest yet  T-duality is not,   in the DFT action given in  the form (\ref{DFTS})  it is quite  the opposite. \\

In order to manifest both the $\ODD$ structure and the gauge symmetry,  in our previous work~\cite{Jeon:2010rw},  we conceived  a differential  geometry characterized by a  \textit{projection}   satisfying the following defining properties, 
\be
\ba{ll}
P_{A}{}^{B}P_{B}{}^{C}=P_{A}{}^{C}\,,~~~~&~~~P_{AB}=P_{BA}\,.
\ea
\label{prodef}
\ee
Further demanding that the  upper left $D\times D$ block of $2P{-1}$ is non-degenerate,  the projection is related to the matrix, $\cH_{AB}$ (\ref{gH}), by
\be
P_{A}{}^{B}=\half(\delta_{A}{}^{B}+\cH_{A}{}^{B})\,.
\label{PcH}
\ee
In terms of a certain   differential operator compatible with the projection -- which we review later --   we were able to identify    the underlying differential geometry  of  the double field theory and, in particular, to  rewrite  the original  DFT action (\ref{DFTS}) in a compact manner,\footnote{Shortly after our work~\cite{Jeon:2010rw},  an alternative approach to  the underlying differential geometry  of  the double field theory was proposed by Hohm and Kwak~\cite{Hohm:2010xe} based on  earlier works by Siegel~\cite{Siegel:1993th,Siegel:1993xq}.  It differs   from our approach, as it  postulates      a covariant derivative whose  connection is not \textit{a priori}  a physical variable of the  double field theory.}
\be
\dis{S_{\scriptstyle{\rm{DFT}}}=\int\rd y^{2D}~e^{-2d}\,\cH^{AB}\left(4\nabla_{A}d\,\nabla_{B}d+S_{AB}\right)\,.}
\label{DFTS2}
\ee

In this paper, we apply our differential  geometric tools in \cite{Jeon:2010rw} to Yang-Mills theory with  an arbitrary gauge group, $\G$.  We construct a Yang-Mills action  which is covariant under the  $\ODD$ rotation and invariant under 
three-types of  gauge transformations: non-Abelian Yang-Mills, diffeomorphism and one-form gauge symmetries.  The latter two amount to  the DFT gauge symmetry, as summarized below:
\[
\ba{l}
\bullet~~~\ODD~\mbox{T-duality}\\
\bullet~~~\mbox{Gauge~symmetry~~~~~~}
\left\{\ba{l}
\mbox{Yang-Mills~gauge~symmetry}\\
\mbox{DFT~gauge~symmetry~~~~~~~~~~~~~}
\left\{\ba{l}
\mbox{Diffeomorphism} \\
\mbox{One-form gauge~symmetry~for~} B_{\mu\nu}
\ea
\right.
\ea
\right.
\ea
\]
In double field formulation,   our action couples a single $\ODD$ vector potential to the  closed string  double field theory (\ref{DFTS2}), keeping  the $\ODD$ T-duality and all the gauge symmetries manifest.  In terms of  undoubled  component fields, the T-duality works in a nontrivial way and  the action couples a   usual  Yang-Mills gauge field, $A_{\mu}$,  to an additional  one-form field, $\phi_{\mu}$,   and also to the closed string background fields which consist of  the  dilaton, graviton and the  two-form gauge field, $\phi$, $g_{\mu\nu}$, $B_{\mu\nu}$.  \\

In section~\ref{SECREVIEW}, we review our previous work~\cite{Jeon:2010rw}  on the differential geometry for  the closed string  double field theory, and set up our notations. In section~\ref{SECYM}, we present  our $\ODD$ covariant  Yang-Mills theory, both in the double field formulation (subsection~\ref{subsecDFF}) and also in terms of undoubled component fields (subsection~\ref{subsecCFF}). We conclude with some comments in section~\ref{SecCON}.\\
%%%
%%\begin{itemize}
%%\item $\ODD$ T-duality
%%\item Gauge symmetry 
%%\begin{enumerate}
%%\item Yang-Mills gauge symmetry
%%\item DFT gauge symmetry
%%\begin{enumerate}
%%\item Diffeomorphism
%%\item One-form gauge symmetry
%%\end{enumerate}
%%\end{enumerate}
%%\end{itemize}
%%%
%%%
%%\[
%%\ba{l}
%%\left\{
%%\ba{l}
%%\mbox{DFT~gauge~symmetry~~~~~~~~~~~}
%%\left\{
%%\ba{l}
%%\mbox{Diffeomorphism} \\
%%\mbox{One-form gauge~symmetry}
%%\ea
%%\right.\\
%%\mbox{Yang-Mills~gauge~symmetry}
%%\ea
%%\right.
%%\\
%%\ODD~\mbox{T-duality}
%%\ea
%%\]
%%%

\newpage

  %%%%%%%%%%%%%%%%%%%%%%%%%%%%%%%%%%%%%%%%%%%%%%%%%%%%%%%%%%%%%%%%%%%%%%%%%%%%%%%%%%%%%%%%%%%%%%%%%%%%%%%%%%%%%%%%%%%%%%%%%%%%%%%%%%%%%%%%%%%%%%%%%%%%%%%%%%%%%%%%%%%%%%%%%%
\section{Differential geometry compatible with a projection: \textit{review}\label{SECREVIEW}}
In double field theory,  the usual definition of Lie derivative is generalized to~\cite{Grana:2008yw,Hohm:2010pp,Jeon:2010rw}
\be
\tcL_{X}\Tw_{A_{1}A_{2}\cdots A_{n}}:= X^{B}\partial_{B}\Tw_{A_{1}A_{2}\cdots A_{n}}+\omega\partial_{B}X^{B}\Tw_{A_{1}A_{2}\cdots A_{n}} +\dis{\sum_{i=1}^{n}}\,2\partial_{[A_{i}}X_{B]}\Tw_{A_{1}\cdots A_{i-1}}{}^{B}{}_{A_{i+1}\cdots A_{n}}\,,
\label{tcLdef}
\ee
where $\omega$ is the weight of each field, $\Tw_{A_{1}A_{2}\cdots A_{n}}$, and $X^{A}$ is a local gauge parameter, of which half  corresponds to the $D$-dimensional diffeomorphism parameter, $\delta x^{\mu}$, and the other half matches the one-form gauge symmetry  parameter, $\Lambda_{\nu}$.  Up to the level matching constraint (\ref{constraint}),  the commutator of them is  closed by  the $\mathbf{c}$-bracket introduced by Siegel~\cite{Siegel:1993th},\footnote{Upon the level matching constraints  the  $\mathbf{c}$-bracket itself reduces to the Courant bracket~\cite{Courant},  as recognized in \cite{Hull:2009zb}.}
\be
\ba{ll}
{}[\tcL_{X},\tcL_{Y}]\equiv\tcL_{[X,Y]_{\rbC}}\,,~~&~~
{}[X,Y]^{A}_{\rbC}=X^{B}\partial_{B}Y^{A}-Y^{B}\partial_{B}X^{A}+\half Y^{B}\partial^{A}X_{B}-\half X^{B}\partial^{A}Y_{B}\,.
%%%
%%[X,Y]_{\rbC}=[X,Y]+\half(Y^{A}\partial^{B}X_{A}-X^{A}\partial^{B}Y_{A})\partial_{B}\,.
%%%
\label{Courant}
\ea
\ee
By definition  in double field theory, \textit{covariant}  {tensors} (${\omega=0}$) or {tensor densities} follow the 
gauge transformation rule  dictated  by the generalized Lie derivative,
\be
\delta_{X}\Tw_{A_{1}A_{2}\cdots A_{n}}=\tcL_{X}\Tw_{A_{1}A_{2}\cdots A_{n}}\,.
\ee
Examples include   for a tensor, $\cH_{AB}$, and  for a scalar density with weight one, $e^{-2d}$, such that\footnote{Another example of a  covariant tensor  is the  $\mathbf{c}$-bracket of two covariant vectors, $\delta_{X}\left([X,Y]^{A}_{\rbC}\right)\equiv\tcL_{X}\!\left([X,Y]^{A}_{\rbC}\right)$ \cite{Gualtieri:2003dx}. }
\be
\ba{cll}
\delta_{X}\cH_{AB}\!\!&=\tcL_{X}\cH_{AB}\!\!&\!\!=X^{C}\partial_{C}\cH_{AB}+(\partial_{A}X_{C}-\partial_{C}X_{A})\cH^{C}{}_{B}+
(\partial_{B}X_{C}-\partial_{C}X_{B})\cH_{A}{}^{C}\,,\\
\delta_{X}\left(e^{-2d}\right)\!\!&=\tcL_{X}\left(e^{-2d}\right)\!\!
&\!\!=\partial_{A}\left(X^{A}e^{-2d}\right)\,.
\label{cHdTr}
\ea
\ee
The latter suggests,  with  $\tcL_{X}\left(e^{-2d}\right)=-2{(\tcL_{X} d)}e^{-2d}$,
\be
\delta_{X}d=\tcL_{X} d:=X^{A}\partial_{A}d -\half\partial_{B}X^{B}\,.
\label{tcLdil}
\ee
The  DFT action (\ref{DFTS}) is indeed invariant under the above  gauge transformation (\ref{cHdTr}), as first shown in \cite{Hohm:2010pp}.  \\

In our previous work~\cite{Jeon:2010rw},  we introduced the following    \textit{projection-compatible derivative},  $\nabla_{C}$, which acts  on tensors,  tensor densities as well as their  
derivative-descendants as
\be
\nabla_{C}\Tw_{A_{1}A_{2}\cdots A_{n}}
=\partial_{C}\Tw_{A_{1}A_{2}\cdots A_{n}}-\omega\Gamma^{B}{}_{BC}\Tw_{A_{1}A_{2}\cdots A_{n}}+
\sum_{i=1}^{n}\,\Gamma_{CA_{i}}{}^{B}\Tw_{A_{1}\cdots A_{i-1}BA_{i+1}\cdots A_{n}}\,,
\label{DdefwT}
\ee
where the connection is, with the projection, (\ref{prodef}),  (\ref{PcH}), and its complementary  projection,  
$\bP:={1-P}$,  given by
\be
\Gamma_{CAB}:=2P_{[A}{}^{D}{\bP}_{B]}{}^{E}\partial_{C}P_{DE}+2\left({\bP}_{[A}{}^{D}{\bP}_{B]}{}^{E}-P_{[A}{}^{D}P_{B]}{}^{E}\right)\partial_{D}P_{EC}\,.
\label{connectionG}
\ee
This  connection  was uniquely determined in terms of the projections and their derivatives,\footnote{One possible generalization of (\ref{connectionG}) which we have not taken seriously is to include the dilaton and its derivative in the connection,
\[
\Gamma_{CAB}\rightarrow\Gamma^{\prime}_{CAB}:=\Gamma_{CAB}-\textstyle{\frac{2}{D{-1}}}(
P_{CA}P_{BD}-P_{CB}P_{AD}+\bP_{CA}\bP_{BD}-\bP_{CB}\bP_{AD})\nabla^{D}d\,.
\]
The resulting derivative  satisfies (\ref{pseudoconstant2}), (\ref{sympropG}), (\ref{CourantD}) and further that $\na^{\prime}d=\partial_{A}d+\half\Gamma^{\prime B}{}_{BA}=0$, whilst  it does not affect the covariant quantities in  (\ref{PDPT}).  However, it becomes singular in the case of  ${D=1}$.}  by requiring  
\be
\ba{ll}
\nabla_{A}\eta_{BC}=0\,,~~~~~~~~~~~~&~~~~~~~~\nabla_{A}P_{BC}=0\,,~~~~~~~~~~~~~~~~~~~~~~~~~~~
\ea
\label{pseudoconstant2}
\ee
and
\be
\ba{ll}
\Gamma_{CAB}+\Gamma_{CBA}=0\,,~~~~&~~~~\Gamma_{ABC}+\Gamma_{CAB}+\Gamma_{BCA}=0\,.
\ea
\label{sympropG}
\ee
The  unique   feature of the  projection-compatible  derivative is that, acting on a covariant tensor, although it does not lead to a covariant quantity,
\be
\left(\delta_{X}-\tcL_{X}\right)\nabla_{C}T_{A_{1}A_{2}\cdots A_{n}}\equiv
2\sum_{i=1}^{n}\left(P_{A_{i}}{}^{D}P_{B}{}^{E}P_{C}{}^{F}+{\bP}_{A_{i}}{}^{D}{\bP}_{B}{}^{E}{\bP}_{C}{}^{F}\right)
\partial_{F}\partial_{[D}X_{E]}T_{A_{1}\cdots A_{i-1}}{}^{B}{}_{A_{i+1}\cdots A_{n}}\,,
\label{DTodd}
\ee
after being  contracted properly with the projections,  it can be covariantized as 
\be
\ba{l}
\Big(\delta_{X}-\tcL_{X}\Big)\Big(P_{C}{}^{D}{\bP}_{A_{1}}{}^{B_{1}}{\bP}_{A_{2}}{}^{B_{2}}\cdots{\bP}_{A_{n}}{}^{B_{n}}
\nabla_{D}T_{B_{1}B_{2}\cdots B_{n}}\Big)\equiv 0\,,\\
\Big(\delta_{X}-\tcL_{X}\Big)\Big({\bP}_{C}{}^{D}P_{A_{1}}{}^{B_{1}}P_{A_{2}}{}^{B_{2}}\cdots P_{A_{n}}{}^{B_{n}}
\nabla_{D}T_{B_{1}B_{2}\cdots B_{n}}\Big)\equiv 0\,.
\ea
\label{PDPT}
\ee
Thanks to the  symmetric properties (\ref{sympropG}),  all the ordinary derivatives  in the definitions  of the generalized Lie derivative (\ref{tcLdef}) and   the $\mathbf{c}$-bracket (\ref{Courant}) can be   replaced by our projection-compatible  derivatives,\footnote{The weight of a gauge symmetry  parameter  is taken to be zero,   such that $\na_{A}X^{B}=\partial_{A}X^{B}+\Gamma_{A}{}^{B}{}_{C}X^{C}$.}
\be
\ba{cll}
\tcL_{X}\Tw_{A_{1}\cdots A_{n}}
&\!=\!&X^{B}\nabla_{B}\Tw_{A_{1}\cdots A_{n}}+\omega\nabla_{B}X^{B}\Tw_{A_{1}\cdots A_{n}} +\sum_{i=1}^{n}2\nabla_{[A_{i}}X_{B]}\Tw_{A_{1}\cdots A_{i-1}}{}^{B}{}_{A_{i+1}\cdots A_{n}}\,,\\
{}[X,Y]^{A}_{\rbC}&\!=\!&X^{B}\na_{B}Y^{A}-Y^{B}\na_{B}X^{A}+\half Y^{B}\na^{A}X_{B}-\half X^{B}\na^{A}Y_{B}\,.
\ea
\label{CourantD}
\ee
Postulating this property to hold also for the gauge transformation of the dilaton (\ref{tcLdil}), and  writing 
\be
\ba{ll}
\nabla_{A}(e^{-2d})=(-2\nabla_{A}d)e^{-2d}\,,~~~~&~~~~ 
\nabla_{A}\nabla_{B}(e^{-2d})=(-2\nabla_{A}\nabla_{B}d+4\nabla_{A}d\na_{B}d)e^{-2d}\,,
\ea
\ee
it is natural  further to set, as if $\nabla_{A}d$ has  trivial weight,
\be
\ba{ll}
\nabla_{A}d:=\partial_{A}d+\half\Gamma^{B}{}_{BA}\,,~~~~~&~~~~\na_{A}\nabla_{B}d:=\partial_{A}\nabla_{B}d+\Gamma_{AB}{}^{C}\nabla_{C}d\,.
\ea
\label{nablad}
\ee

Now, with the  curvature defined   in  standard way,
\be
R_{CDAB}:=\partial_{A}\Gamma_{BCD}-\partial_{B}\Gamma_{ACD}+\Gamma_{AC}{}^{E}\Gamma_{BED}-\Gamma_{BC}{}^{E}\Gamma_{AED}\,,
\label{RABCD}
\ee
if we set
\be
\ba{ll}
S_{ABCD}:=\half\left(R_{ABCD}+R_{CDAB}-\Gamma^{E}{}_{AB}\Gamma_{ECD}\right)\,,~~~~&~~~~
S_{AB}:=S^{C}{}_{ACB}\,,
\ea
\ee
the following quantities are all gauge covariant~\cite{Jeon:2010rw},
\begin{eqnarray}
&&\cR_{AB}:=P_{A}{}^{C}\bP_{B}{}^{D}\left(S_{CD}+2\na_{(C}\nabla_{D)}d\right)\,,\label{cRAB}\\
&&\cR:=\cH^{AB}\left(4\na_{A}\nabla_{B}d-4\nabla_{A}d\,\nabla_{B}d+S_{AB}\right)\,,\label{cR}\\
&&P^{AB}(\na_{A}-2\nabla_{A}d)V_{B}\,,\\
&&\bP^{AB}(\na_{A}-2\nabla_{A}d)V_{B}\,,\\
&&P^{AB}{\bP}_{C_{1}}{}^{D_{1}}\cdots{\bP}_{C_{n}}{}^{D_{n}}
\left[\na_{A}\na_{B}T_{D_{1}\cdots D_{n}}-2(\nabla_{A}d) \na_{B}T_{D_{1}\cdots D_{n}}\right]\,,\\
&&{\bP}^{AB}P_{C_{1}}{}^{D_{1}}\cdots P_{C_{n}}{}^{D_{n}}
\left[\na_{A}\na_{B}T_{D_{1}\cdots D_{n}}-2(\nabla_{A}d)\na_{B}T_{D_{1}\cdots D_{n}}\right]\,,
\end{eqnarray}
in addition to  the ones in (\ref{PDPT}),\footnote{Successive application of (\ref{Tobeused}) with  more than one covariant vectors also leads to the following gauge covariant higher order derivatives: 
\[
\ba{l}\left(\prod_{i=1}^{m}V_{i}^{B}P_{B}{}^{C}\na_{C}\right){\bP}_{A_{1}}{}^{B_{1}}{\bP}_{A_{2}}{}^{B_{2}}\cdots{\bP}_{A_{n}}{}^{B_{n}}T_{B_{1}B_{2}\cdots B_{n}}\,,\\
\left(\prod_{i=1}^{m}V_{i}^{B}\bP_{B}{}^{C}\na_{C}\right)P_{A_{1}}{}^{B_{1}}P_{A_{2}}{}^{B_{2}}\cdots P_{A_{n}}{}^{B_{n}}T_{B_{1}B_{2}\cdots B_{n}}\,.
\ea
\]}
\be
\ba{l}
P_{C}{}^{D}{\bP}_{A_{1}}{}^{B_{1}}{\bP}_{A_{2}}{}^{B_{2}}\cdots{\bP}_{A_{n}}{}^{B_{n}}
\na_{D}T_{B_{1}B_{2}\cdots B_{n}}\,,\\
{\bP}_{C}{}^{D}P_{A_{1}}{}^{B_{1}}P_{A_{2}}{}^{B_{2}}\cdots P_{A_{n}}{}^{B_{n}}
\na_{D}T_{B_{1}B_{2}\cdots B_{n}}\,.
\ea
\label{Tobeused}
\ee
As a matter of fact, up to a surface term, the double field theory Lagrangian in  (\ref{DFTS2}) is equivalent to $e^{-2d}\cR$, while  its equations of motion for the dilaton and the projection are $\cR=0$ and $\cR_{(AB)}=0\,$ respectively. \\

Some useful identities to note are 
\begin{eqnarray}
&S_{ABCD}=S_{[AB][CD]}\,,~~~~~~~~~~~S_{ABCD}=S_{CDAB}\,,~~~~~~~~~~~S_{A[BCD]}=0\,,&\\
&P_{A}{}^{E}\bP_{B}{}^{F}P_{C}{}^{G}\bP_{D}{}^{H}S_{EFGH}\equiv 0\,,~~~~~~~~P_{A}{}^{E}P_{B}{}^{F}\bP_{C}{}^{G}\bP_{D}{}^{H}S_{EFGH}\equiv 0\,,&\\
&4\na_{A}\nabla^{A}d-4\nabla_{A}d\,\nabla^{A}d+S\equiv 0\,.&
\end{eqnarray}
Under an arbitrary infinitesimal transformation of the projection satisfying 
\be
\delta P= P\delta P\bP+\bP\delta P P\,,
\label{infPtr}
\ee
the connection and $S_{ABCD}$ transform as
\be
\ba{ll}
\delta\Gamma_{CAB}=&2P_{[A}{}^{D}\bP_{B]}{}^{E}\na_{C}\delta P_{DE}+2(\bP_{[A}{}^{D}\bP_{B]}{}^{E}-
P_{[A}{}^{D}P_{B]}{}^{E})\na_{D}\delta P_{EC}\\
{}&{}-\Gamma_{FDE\,}\delta(P_{C}{}^{F}P_{A}{}^{D}P_{B}{}^{E}+
\bP_{C}{}^{F}\bP_{A}{}^{D}\bP_{B}{}^{E})\,,\\
\multicolumn{2}{l}{\delta S_{ABCD}=\na_{[A}\delta\Gamma_{B]CD}+\na_{[C}\delta\Gamma_{D]AB}\,.}
\ea
\label{infGStr}
\ee
~\\~\\

%%%%%%%%%%%%%%%%%%%%%%%%%%%%%%%%%%%%%%%%%%%%%%%%%%%%%%%%%%%%%%%%%%%%%%%%%%%%%%%%%%%%%%%%%%%%%%%%%%%%%%%%%%%%%%%%%%%%%%%%%%%%%%%%%%%%%%%%%%%%%%%%%%%%%%%%%%%%%%%%%%%%%%%%%%
\section{$\ODD$ covariant  Yang-Mills theory\label{SECYM}}
\subsection{Double field formulation\label{subsecDFF}}
Our main result in  the present paper comes from generalizing  the previous analysis  on the covariant quantities, especially  (\ref{Tobeused}), to Yang-Mills theory with a generic  non-Abelian gauge  group, $\G$.  We postulate a  DFT  vector potential, $V_{A}$, which is in the adjoint   representation of the Lie algebra of the gauge group, $\cG$.  For a DFT tensor, $T_{A_{1}A_{2}\cdots A_{n}}$ which is  in the fundamental representation of $\cG$, we  define  with the projection-compatible derivative (\ref{DdefwT}),
\be
\cD_{C}T_{A_{1}A_{2}\cdots A_{n}}:=\na_{C}T_{A_{1}A_{2}\cdots A_{n}}-iV_{C}T_{A_{1}A_{2}\cdots A_{n}}\,.
\label{defcD}
\ee
This derivative is covariant with respect to the usual Yang-Mills gauge symmetry: with $\mbg\in\G$, under
\be
\ba{cll}
T_{A_{1}A_{2}\cdots A_{n}}&\longrightarrow& ~\mbg T_{A_{1}A_{2}\cdots A_{n}}\,,\\
V_{A}&\longrightarrow& \mbg V_{A}\mbg^{-1}-i(\partial_{A}\mbg)\mbg^{-1}\,,
\ea
\ee
the derivative transforms in a  covariant fashion,
\be
\cD_{C}T_{A_{1}A_{2}\cdots A_{n}}~\longrightarrow~\mbg\cD_{C}T_{A_{1}A_{2}\cdots A_{n}}\,.
\ee
Note that the projection and the dilaton are all Yang-Mills gauge singlets such that  the projection-compatible derivative (\ref{DdefwT}) does not change under the  Yang-Mills gauge transformation. 

The commutator of the above derivatives   reads
\be
{}[\cD_{A},\cD_{B}]T_{C_{1}C_{2}\cdots C_{n}}=-iF_{AB}T_{C_{1}C_{2}\cdots C_{n}}
-\Gamma^{D}{}_{AB}\cD_{D}T_{C_{1}C_{2}\cdots C_{n}}+
\sum_{i=1}^{n}R_{C_{i}DAB}\,T_{C_{1}\cdots C_{i-1}}{}^{D}{}_{C_{i+1}\cdots C_{n}}\,,
\label{COMM}
\ee
where $R_{CDAB}$ is  the curvature given in (\ref{RABCD}), and $F_{AB}$ is the   field strength of the vector potential,
\be
F_{AB}=\partial_{A}V_{B}-\partial_{B}V_{A}-i\left[V_{A},V_{B}\right]\,,
\label{FAB}
\ee
which surely  transforms covariantly under the Yang-Mills gauge  transformation,
\be
F_{AB}\,~\longrightarrow~\,\mbg F_{AB} \mbg^{-1}\,.
\ee
However, this field strength is not  DFT gauge covariant, 
\be
\delta_{X}F_{AB}\neq\tcL_{X}F_{AB}\,.
\ee
It is necessary to  utilize  the projection compatible derivative as in  (\ref{Tobeused}). Hence,  instead of (\ref{FAB}) we consider
\be
\cF_{AB}:=\na_{A}V_{B}-\na_{B}V_{A}-i\left[V_{A},V_{B}\right]=F_{AB}-\Gamma^{C}{}_{AB}V_{C}\,.
\ee
Although it is not covariant under the Yang-Mills gauge symmetry,
\be
\cF_{AB}\,~\longrightarrow~\,\mbg\cF_{AB} \mbg^{-1}+i\Gamma^{C}{}_{AB}(\partial_{C}\mbg)\mbg^{-1}\,,
\ee
 when its two $\ODD$ vector indices are projected into  opposite chiralities, 
\be
P_{A}{}^{C}\bP_{B}{}^{D}\cF_{CD}\,,
\label{FULLCOV}
\ee
it becomes covariant with respect to both the Yang-Mills and the DFT gauge symmetries, thanks to the level matching constraint (\ref{constraint}) imposed on   the explicit expression of the connection (\ref{connectionG}), 
\be
\ba{ccc}
P_{A}{}^{C}\bP_{B}{}^{D}\cF_{CD}&\longrightarrow& P_{A}{}^{C}\bP_{B}{}^{D}\mbg\cF_{CD} \mbg^{-1}\,,\\
\delta_{X}(P_{A}{}^{C}\bP_{B}{}^{D}\cF_{CD})&\equiv&\tcL_{X}(P_{A}{}^{C}\bP_{B}{}^{D}\cF_{CD})\,.
\ea
\ee
Therefore,   our double field  formulation of a Yang-Mills action is
\be
S_{\scriptstyle{\rm{YM}}} =\gYM^{-2}\dis{\int\rd y^{2D}~e^{-2d\,}\Tr\!\left(P^{AB}\bP^{CD}\cF_{AC}\cF_{BD}\right)\,,}
\label{SYM}
\ee
which can be coupled to the  closed string DFT  (\ref{DFTS2}) as
\be
S_{\scriptstyle{\rm{DFT}}}+S_{\scriptstyle{\rm{YM}}}=\int\rd y^{2D}~e^{-2d}\left[{
\cH^{AB}\left(4\nabla_{A}d\,\nabla_{B}d+S_{AB}\right)+\gYM^{-2}\,\Tr\!\left(P^{AB}\bP^{CD}\cF_{AC}\cF_{BD}\right)}\right]\,.
\ee
These actions are manifestly  $\ODD$ covariant,  and  invariant under both  the Yang-Mills and the DFT gauge transformations.\\

%%%%%%%%%%%%%%%%%%%%%%%%%%%%%%%%%%%%%%%%%%%%%%%%%%%%%%%%%%%%%%%%%%%%%%%%%%%%%%%%%%%%%%%%%%%%%%%%%%%%%%%%%%%%%%%%%%%%%%%%%%%%%%%%%%%%%%%%%%%%%%%%%%%%%%%%%%%%%%%%%%%%%%%%%%
\subsection{Component field formulation\label{subsecCFF}}
Here we rewrite  the above   double field  formulation of  a Yang-Mills action (\ref{SYM}) in terms of ordinary  undoubled $D$-dimensional component fields, in a similar fashion  that  the closed string DFT  action, $S_{\scriptstyle{\rm{DFT}}}$  (\ref{DFTS2}), reduces to the more familiar looking effective action,  $S_{\eff}$ (\ref{NSaction}), upon the level matching  constraint.\\

We first decompose the DFT vector potential into a chiral and an anti-chiral  vectors,
\be
\ba{lll}
V_{A}=V^{+}_{A}+V^{-}_{A}\,,~~~~&~~~~V^{+}_{A}=P_{A}{}^{B}V_{B}\,,~~~~&~~~~V^{-}_{A}=\bP_{A}{}^{B}V_{B}\,,
\ea
\ee
such that $\cH_{A}{}^{B}V^{\pm}_{B}=\pm V^{\pm}_{A}$. The chiral and anti-chiral vectors    assume  the following  generic forms,  
\be
\ba{ll}
V^{+}_{A}=\half\left(\ba{c}A^{+\lambda}\\ (g{+B})_{\mu\nu}A^{+\nu}\ea\right)\,,~~~~&~~~~
V^{-}_{A}=\half\left(\ba{c}-A^{-\lambda}\\(g{-B})_{\mu\nu} A^{-\nu}\ea\right)\,.
\ea
\ee
With the field redefinition, 
\be
\ba{ll}
A_{\mu}:=\half(A^{+}_{\mu}+A^{-}_{\mu})\,,~~~~&~~~~
\phi_{\mu}:=\half(A^{+}_{\mu}-A^{-}_{\mu})\,,
\ea
\ee
which is equivalent to  $A^{\pm}_{\mu}=A_{\mu}\pm\phi_{\mu}$, the DFT vector potential can be parametrized  by
\be
V_{A}=\left(\ba{c}\phi^{\lambda}\\A_{\mu}+B_{\mu\nu}\phi^{\nu}\ea\right)\,.
\ee
Note that the $D$-dimensional vector indices, $\mu,\nu$, are here and henceforth  freely  raised or lowered by the $D$-dimensional metric, $g_{\mu\nu}$, in  the usual  manner.\\

Direct computation shows, turning off the $\tx$-dependence, 
\be
P_{A}{}^{C}\bP_{B}{}^{D}\cF_{CD}\equiv\textstyle{\frac{1}{4}}\left(\ba{cc}
-\hat{f}^{\lambda\mu}~&~\hat{f}^{\lambda\tau}(g{+B})_{\tau\nu}\\
-(g{+B})_{\rho\sigma}\hat{f}^{\sigma\mu}~&~(g{+B})_{\rho\sigma}\hat{f}^{\sigma\tau}(g{+B})_{\tau\nu}
\ea\right)\,,
\label{TbT}
\ee
where  we set 
\be
\ba{cll}
\hat{f}_{\mu\nu}\!\!&:=&\!f_{\mu\nu}-D_{\mu}\phi_{\nu}-D_{\nu}\phi_{\mu}+i\left[\phi_{\mu},\phi_{\nu}\right]+H_{\mu\nu\lambda}\phi^{\lambda}\,,\\
D_{\mu}\phi_{\nu}\!\!&:=&\!\na_{\mu}\phi_{\nu}-i\left[A_{\mu},\phi_{\nu}\right]=\partial_{\mu}\phi_{\nu}-\phi_{\lambda}\gamma_{\mu\nu}^{~\lambda}-i\left[A_{\mu},\phi_{\nu}\right]\,,\\
f_{\mu\nu}\!\!&:=&\!\partial_{\mu}A_{\nu}-\partial_{\nu}A_{\mu}-i\left[A_{\mu},A_{\nu}\right]\,,\\
H_{\lambda\mu\nu}\!\!&:=&\!\partial_{\lambda}B_{\mu\nu}+\partial_{\mu}B_{\nu\lambda}+\partial_{\nu}B_{\lambda\mu}\,.
\ea
\ee
Unlike (\ref{DdefwT}) and (\ref{defcD}), in our $D$-dimensional notation, $\na_{\mu}$ denotes the usual diffeomorphism covariant derivative involving  the  Christoffel symbol, $\gamma_{\mu\nu}^{~\lambda}=\half g^{\lambda\rho}(\partial_{\mu}g_{\rho\nu}+\partial_{\nu}g_{\mu\rho}-\partial_{\rho}g_{\mu\nu})$, and $D_{\mu}$  is the diffeomorphism plus Yang-Mills gauge covariant derivative.\\

It is worth while to note 
\be
\ba{cll}
\hat{f}_{\mu\nu}\!\!&=&\!\!\na_{\mu}A^{-}_{\nu}-\na_{\nu}A^{+}_{\mu}-i\left[A^{+}_{\mu},A^{-}_{\nu}\right]+H_{\mu\nu\lambda}\phi^{\lambda}\,,\\
\hat{f}_{[\mu\nu]}\!\!&=&\!\!f_{\mu\nu}+i\left[\phi_{\mu},\phi_{\nu}\right]+H_{\mu\nu\lambda}\phi^{\lambda}\,,\\
\hat{f}_{(\mu\nu)}\!\!&=&\!\!-(D_{\mu}\phi_{\nu}+D_{\nu}\phi_{\mu})\,,
\ea
\ee
and for (\ref{TbT})
\be
P_{A}{}^{C}\bP_{B}{}^{D}\cF_{CD}\equiv P_{A}{}^{C}\bP_{B}{}^{D}
\left(\ba{ll}0&~0\\
0&\hat{f}_{\mu\nu}\ea\right)_{CD}\,.
\ee
Now, from (\ref{TbT}), it is straightforward to show that the Yang-Mills action in the   double field  formulation (\ref{SYM}) reduces to
\be
S_{\scriptstyle{\rm{YM}}} \equiv\gYM^{-2}\dis{\int\rd x^{D}~\sqrt{-g}e^{-2\phi}\,\Tr\!\left(-\textstyle{\frac{1}{4}}\hat{f}^{\mu\nu}\hat{f}_{\mu\nu}\right)\,,}
\label{SYM2}
\ee
and hence,
\be
S_{\rm{DFT}}+S_{\scriptstyle{\rm{YM}}}\equiv
\int\rd x^{D}\sqrt{-g}e^{-2\phi}\left[\,R_{g}+4(\partial \phi)^{2}-\textstyle{\frac{1}{12}}H^{2}-\textstyle{\frac{1}{4}}\gYM^{-2}\,
\Tr\!\left(\hat{f}{}^{\,2}\right)\right]\,.
\label{SDFTYM}
\ee
Explicitly, we have for  $S_{\scriptstyle{\rm{YM}}}$ (\ref{SYM2}),
\be
\ba{ll}
\Tr\Big(\hat{f}_{\mu\nu}\hat{f}^{\mu\nu}\Big)=
\Tr\Big(&\!\!\! f_{\mu\nu}f^{\mu\nu}+2 D_{\mu}\phi_{\nu}D^{\mu}\phi^{\nu}+2D_{\mu}\phi_{\nu}D^{\nu}\phi^{\mu}-
[\phi_{\mu},\phi_{\nu}][\phi^{\mu},\phi^{\nu}]\\
{}&+2if_{\mu\nu}[\phi^{\mu},\phi^{\nu}]+2\left(f^{\mu\nu}+i[\phi^{\mu},\phi^{\nu}]
\right)H_{\mu\nu\sigma}\phi^{\sigma}+H_{\mu\nu\sigma}H^{\mu\nu}{}_{\tau}\phi^{\sigma}\phi^{\tau}~\Big)\,.
\ea
\ee
The above actions (\ref{SYM2}), (\ref{SDFTYM}) are clearly   invariant under both  the Yang-Mills and the DFT gauge symmetries. Moreover, though not manifest,  by construction  it enjoys  T-duality. \\~\\~\\\newpage

%%%%%%%%%%%%%%%%%%%%%%%%%%%%%%%%%%%%%%%%%%%%%%%%%%%%%%%%%%%%%%%%%%%%%%%%%%%%%%%%%%%%%%%%%%%%%%%%%%%%%%%%%%%%%%%%%%%%%%%%%%%%%%%%%%%%%%%%%%%%%%%%%%%%%%%%%%%%%%%%%%%%%%%%%%

%%%%%%%%%%%%%%%%%%%%%%%%%%%%%%%%%%%%%%%%%%%%%%%%%%%%%%%%%%%%%%%%%%%%%%%%%%%%%%%%%%%%%%%%%%%%%%%%%%%%%%%%%%%%%%%%%%%%%%%%%%%%%%%%%%%%%%%%%%%%%%%%%%%%%%%%%%%%%%%%%%%%%%%%%%
\section{Comments\label{SecCON}}
We recall the  DFT tensor (\ref{FULLCOV}) which is fully covariant under the   $\ODD$ T-duality  as well as all the  gauge symmetries,  to set
\be
\hcF_{AB}:=P_{A}{}^{C}\bP_{B}{}^{D}\cF_{CD}\,.
\ee
Apart from $\Tr(\hcF^{AB}\hcF_{AB})$ which essentially leads to our DFT formulation of the Yang-Mills  action (\ref{SYM}),  the following quantity of even power in the field strength  is also  fully  covariant,
\be
\Tr\left(\hcF^{A_{1}B_{1}}\hcF_{A_{2}B_{1}}\hcF^{A_{2}B_{2}}\hcF_{A_{3}B_{2}}\cdots\hcF^{A_{n}B_{n}}\hcF_{A_{1}B_{n}}\right)\,.
\ee
Due to the chirality of $\hcF_{AB}$, there is  no covariant  scalar   with odd power. Especially, for the Abelian group,\footnote{Generalization to   non-Abelian Born-Infeld action is also doable     following various   prescriptions, \textit{e.g.~}\cite{Hagiwara:1981my,Argyres:1989qr,Tseytlin:1997csa,Park:1999gd,Serie:2003nf,Serie:2004wj}.} $\G=\mbox{U}(1)$, we obtain another covariant quantity,\footnote{On the other hand, due to the chirality of $\hcF_{AB}$,  $\,\det(\eta_{AB}+\kappa\hcF_{AB})$ is trivial.}
\be
\det\!\left(\eta_{AB}+\kappa\,\hcF_{AC}\hcF_{B}{}^{C}\right)=
\det\!\left(\eta_{AB}+\kappa\,\hcF_{CA}\hcF^{C}{}_{B}\right)\,,
\ee
where $\kappa$ is a constant and  the determinant is taken over the $\ODD$ vector indices, $A,B$. Since this is a scalar rather than a scalar density, there appears  no compulsory   reason to take a square root   of the determinant constructing   a  Born-Infeld type  action.\\

In the presence of a curved  $D$-brane, string theory can  force a topological  twisting on a usual Yang-Mills theory,    converting scalars into one-form~\cite{Bershadsky:1995qy}.   Especially, when a pure Yang-Mills theory in $(D+D)$-dimensions is reduced to $D$-dimensions, the Lorentz symmetry group  coincides with the $R$-symmetry group. If we diagonalize these two, as in topological twisting theories~\cite{Witten:1988ze,Vafa:1994tf,Yamron:1988qc,Marcus:1995mq,Park:2006kt},  we may obtain the following maximally twisted action,
\be
S_{\scriptstyle{\rm{twisted}}} \equiv-\gYM^{-2}\dis{\int\rd x^{D}~\sqrt{-g}\,\Tr\!\left({
\textstyle{\frac{1}{4}}f_{\mu\nu}f^{\mu\nu}
+\half D_{\mu}\phi_{\nu}D^{\mu}\phi^{\nu}-\textstyle{\frac{1}{4}}
[\phi_{\mu},\phi_{\nu}][\phi^{\mu},\phi^{\nu}]+\half R_{\mu\nu}\phi^{\mu}\phi^{\nu}}
\right)\,.}
\label{5D2}
\ee
Intriguingly this twisted action resembles our Yang-Mills action (\ref{SYM2}), although  they differ in some details.\footnote{To confirm the difference, it is necessary to use the identity,
\[
[D_{\mu},D_{\nu}]\phi^{\nu}+R_{\mu\nu}\phi^{\nu}+i\left[f_{\mu\nu},\phi^{\nu}\right]=0\,.
\]}
More precise string theory  interpretation   of our double field formulation of Yang-Mills theory    is desirable (for some related works we refer \cite{Bergshoeff:1994dg,Bergshoeff:1995cg,Chalmers:1997sg}).  Doubled  sigma-model formalism~\cite{Hull:2004in,Hull:2006va,Berman:2007xn,Berman:2007yf} may provide useful  insights.   \newpage

~\\~\\~\\
~\\
~\\
~\\
~\\
\textit{Note added}: After submitting the first version  of this manuscript to arXiv, a related  work  by Hohm and Kwak appeared~\cite{Hohm:2011ex}. Their paper attempts  the double field theory formulation of the heterotic string effective action, and hence the inclusion of Yang-Mills  theories.  It is based on an  enlarged,  yet broken,  $\mathbf{O}(D,D+n)$ T-duality, which differs from ours, \textit{i.e.~}unbroken $\ODD$. \\
~\\
~\\
\noindent\textbf{Acknowledgements}\\
We are grateful to  Seung Ki Kwak  for useful  discussion which motivated our work.  The work was supported by the National Research Foundation of Korea(NRF) grants  funded by the Korea government(MEST) with the grant numbers 2005-0049409 (CQUeST)  and 2010-0002980.
\newpage

%%%%%%%%%%%%%%%%%%%%%%%%%%%%%%%%%%%%%%%%%%%%%%%%%%%%%%%%%%%%%%%%%%%%%%%%%%%%%%%%%%%%%%%%%
%%%%%%%%%%%%%%%%%%%%%%%%%%%%%%%%%%%%%%%%%%%%%%%%%%%%%%%%%%%%%%%%%%%%%%%%%%%%%%%%%%%%%%%%%
%%
%%%
%%\appendix
%%%
%%%%%%%%%%%%%%%%%%%%%%%%%%%%%%%%%%%%%%%%%%%%%%%%%%%%%%%%%%%%%%%%%%%%%%%%%%%%%%%%%%%%%%%%%%%%%%%%%%%%%%%%%%%%%%%%%%%%%%%%%%%%%%%%%%%%%%%%%%%%%%%%%%%%%%%%%%%%%%%%%%%%%%%%%%%%
%%\section{Useful relations\label{SECAPPENDIXuseful}}
%%Here we write some useful identities. \newpage
%%%
%%%%%%%%%%%%%%%%%%%%%%%%%%%%%%%%%%%%%%%%%%%%%%%%%%%%%%%%%%%%%%%%%%%%%%%%%%%%%%%%%%%%%%%%%%%%%%%%%%%%%%%%%%%%%%%%%%%%%%%%%%%%%%%%%%%%%%%%%%%%%%%%%%%%%%%%%%%%%%%%%%%%%%%%%%

\end{document}